\documentclass[12pt]{article}
\usepackage{amssymb}
\usepackage{amsmath}
\usepackage{amstext}
\usepackage{graphicx,epsfig}
\usepackage{epsfig}
\usepackage{verbatim} 
\usepackage{fancyhdr}
\usepackage{fancybox}
\usepackage{color}
\usepackage{ulem,bbold}
\usepackage{enumitem}
\usepackage{subfigure}
\usepackage{bbm}
\usepackage{parskip}
\usepackage{cite}

\newcommand{\Comment}[1]{{}}
\definecolor{MyDarkBlue}{rgb}{0.15,0.15,0.45}
\usepackage[linktocpage=true]{hyperref}
\hypersetup{
colorlinks=true,
citecolor=MyDarkBlue,
linkcolor=MyDarkBlue,
urlcolor=MyDarkBlue,
}

\newcommand{\be}{\begin{equation}}
\newcommand{\ee}{\end{equation}}
\newcommand{\bea}{\begin{eqnarray}}
\newcommand{\eea}{\end{eqnarray}}
\newcommand{\beas}{\begin{eqnarray*}}
\newcommand{\eeas}{\end{eqnarray*}}
\newcommand{\nn}{\nonumber}

\newcommand{\rd}{{\rm d}}

\newcommand{\half}{\frac{1}{2}}

\numberwithin{equation}{section}

\begin{document}


\begin{center}
{\Large \bf{Equivalence of EFTs \\}}
\vspace{.2cm}
{\Large \bf{and Time-Like Extra Dimensions}}
\end{center} 
 \vspace{1truecm}
\thispagestyle{empty} \centerline{
{\large Kurt Hinterbichler}$^{}$\footnote{E-mail: \Comment{\href{mailto:kurt.hinterbichler@case.edu}}{\tt kurt.hinterbichler@case.edu}},\ \ {\large Samanta Saha}$^{}$\footnote{E-mail: \Comment{\href{mailto:sxs2638@case.edu}}{\tt sxs2638@case.edu}} 
}

\vspace{1cm}

\centerline{{\it ${}^{}$CERCA, Department of Physics,}}
\centerline{{\it Case Western Reserve University, 10900 Euclid Ave, Cleveland, OH 44106}} 

 \vspace{1cm}

\begin{abstract} 

A flat 3-brane probing a five dimensional anti-de Sitter space has the same symmetries and symmetry breaking pattern as one probing a five dimensional de Sitter space with two time directions.  Despite the seemingly different physical set ups, we show that the effective field theory of the brane bending mode is identical in both cases.  In particular, despite ``wrong'' signs in the DBI action for the two-time de Sitter case, the theories have the same $S$-matrix.  This is consistent with the expectation that effective field theories are determined solely by their degrees of freedom and pattern of symmetry breaking, even in the case of spacetime symmetries.  We comment further on the equivalence between the Weyl/dilaton and DBI representations of the EFT of broken conformal symmetry.

\end{abstract}

\newpage

\thispagestyle{empty}
\tableofcontents

\setcounter{page}{1}
\setcounter{footnote}{0}

\parskip=5pt
\normalsize

\section{Introduction}

Effective field theory (EFT) is a tool with which we can efficiently describe physics at low energies and long distances, even without detailed knowledge of a complete underlying short distance theory.  It is generally believed that the low energy dynamics of a system, and hence the structure of its effective field theory, is determined solely by its light degrees of freedom, their symmetry group $G$, and the pattern of spontaneous symmetry breaking by which the group is broken to some subgroup $H\subset G$.

In the simplest case where $G$ is an internal symmetry group in a relativistic theory, i.e. $G$ commutes with the Poincare symmetries, Goldstone's theorem guarantees that there will be a massless scalar degree of freedom for each broken symmetry generator \cite{Goldstone:1961eq,Goldstone:1962es}.  These Goldstone scalars parametrize the coset $G/H$ and, since they are massless, are always present in the effective field theory.  The fact that the effective theory is determined by the degrees of freedom and the symmetry breaking pattern is implicit in the coset construction \cite{Coleman:1969sm,Callan:1969sn,Volkov:1973vd} (see \cite{xthschool,Zumino:1970tu} for reviews), via which the invariant building blocks of the effective theory of the Goldstones, and the covariant derivatives through which they couple to other modes, are constructed solely from knowledge of the symmetry group and its breaking pattern.

In the case where the broken symmetries involve spacetime symmetries, the situation is more subtle.  For example, the number of Goldstone modes is in general less than the number of broken generators, and the coset construction has to be supplemented by the addition of inverse Higgs constraints to eliminate the unphysical Goldstones \cite{xthschool,Volkov:1973vd,Ivanov:1975zq,Nielsen:1975hm,Low:2001bw, Bellucci:2002ji,McArthur:2010zm,Watanabe:2012hr,Hidaka:2012ym,Brauner:2014aha}.  In this paper, as a test of whether the symmetry breaking pattern determines the low energy theory in this more general situation, we will study two examples of spacetime symmetry breaking which are seemingly physically distinct, but in which the pattern of symmetry breaking is the same.

Consider the effective theory describing the small fluctuations of a flat Lorentzian 3-brane embedded into a fixed flat 5-dimensional Lorentzian bulk, so that there is one spacelike extra dimension transverse to the brane.  The presence of the brane spontaneously breaks the five dimensional isometry algebra down to the four dimensional isometry algebra, ${iso}(1,4)\rightarrow {iso}(1,3)$.   There is a single Goldstone scalar, $\phi$, which parametrizes the fluctuations into the extra dimension, and the lowest order part of the EFT governing its dynamics is the Dirac-Born-Infeld (DBI) Lagrangian \cite{Born:1934gh,Dirac:1962iy},
\be {\cal L}=- { \Lambda^4} \sqrt{1+{(\partial\phi)^2\over \Lambda^4}} = const.-{1\over 2}(\partial\phi)^2 +{1\over 8 \Lambda^4}(\partial\phi)^4+\cdots\, ,\label{rightsigndbiee}\ee
where $\Lambda$ is a scale set by the tension of the brane.  This is called the ``correct sign'' DBI model, because the ${\cal O}\left(p^4\right)$ part of the 4-point tree amplitude, determined by the $(\partial\phi)^4$ term in \eqref{rightsigndbiee}, is positive,
\be {\cal A}_4= {1\over 4 \Lambda^4}\left(s^2+t^2+u^2\right),\ee
consistent with positivity bounds that follow from assuming a standard UV completion \cite{Adams:2006sv}.

Suppose instead that the extra dimension is timelike, so that we have a Lorentzian 3-brane embedded into a flat 5-dimensional Lorentzian bulk with two time directions.  In this case, the fluctuations are described by
\be {\cal L}= { \Lambda^4} \sqrt{1-{(\partial\phi)^2\over \Lambda^4}} = const.-{1\over 2}(\partial\phi)^2 -{1\over 8 \Lambda^4}(\partial\phi)^4+\cdots\, ,\label{rightsigndbiee2}\ee
which has a different sign under the square root.   This is the ``wrong sign'' DBI theory; by adjusting the overall sign of the Lagrangian we have kept the correct sign for the kinetic term, but now the $(\partial\phi)^4$ term has a different sign and we get a negative ${\cal O}\left(p^4\right)$ 4-point tree amplitude,
\be {\cal A}_4= -{1\over 4 \Lambda^4}\left(s^2+t^2+u^2\right),\label{worgsigneampe}\ee
which is not consistent with the positivity bounds of \cite{Adams:2006sv}.  Because the amplitude is different from that of the correct sign theory, the theory is physically different, as is to be expected because a space with two time dimensions seems to be physically quite different and, some would say, pathological; in fact it is tempting to interpret the wrong-sign amplitude \eqref{worgsigneampe} as a sign of such pathology.  From the symmetry point of view, the physical difference between the two theories is also not surprising; the higher dimensional space with two time directions has a different isometry algebra, ${\frak{iso}}(2,3)$, so the symmetry breaking pattern that we would like to say characterizes the EFT is now ${\frak{iso}}(2,3)\rightarrow {\frak{iso}}(1,3)$, different from that of the correct sign theory.

Now consider a similar setup but let the flat brane be embedded into a bulk five dimensional Lorentzian maximally symmetric space of constant negative curvature, i.e. anti-de Sitter space (AdS).  We denote the bulk space as $AdS_{1,4}$, to indicate that it is a standard AdS space with one time and 4 space dimensions.  Its isometry algebra is $\frak{so}(2,4)$, so the symmetry breaking pattern is $\frak{so}(2,4)\rightarrow \frak{iso}(1,3)$.  The DBI Lagrangian in this case reads
\be {\cal L}= {1\over L^4} e^{4L\phi} \left(1-\sqrt{1+e^{-2L\phi}L^4 (\partial\phi)^2}\right)= -{1\over 2}(\partial\phi)^2 +{L^4\over 8}(\partial\phi)^4+{\rm on\ shell\ trivial}+\cdots \,, \label{introadsede}\ee 
where $L$ is the radius of the bulk AdS space (which we have set to the same scale as the brane tension for simplicity).  The sign of the ${\cal O}\left(p^4\right)$ part of the 4-point tree amplitude, determined by the sign of the $(\partial\phi)^4$ term, is again positive.

Next, in analogy with the 2-time example above, consider embedding the flat 3-brane into a maximally symmetric space of constant positive curvature, but with two time directions, i.e. a 2-time de Sitter (dS) space, denoted $dS_{2,3}$.
The DBI Lagrangian now has a minus sign under the square root,
\be {\cal L}=- {1\over L^4} e^{4L\phi} \left(1-\sqrt{1-e^{-2L\phi}L^4 (\partial\phi)^2}\right)= -{1\over 2}(\partial\phi)^2 -{L^4\over 8}(\partial\phi)^4+{\rm on\ shell\ trivial}+\cdots\, ,  \label{introdseede}\ee 
and the ${\cal O}\left(p^4\right)$ part of the 4-point tree amplitude is now negative.  As in the flat bulk case, from a physical point of view a 2-time de Sitter space seems quite different from, and more pathological than, a standard AdS space, so this different sign in the amplitude might be expected.  

However, from the symmetry point of view we encounter a puzzle: the 2-time de Sitter space has exactly the same symmetry algebra as the standard AdS space, namely $\frak{so}(2,4)$.   This can be seen by realizing $AdS_{1,4}$ as the surface $\eta_{AB}Y^AY^B=-L^2$ embedded into an auxiliary ambient six-dimensional 2-time Minkowski space with Cartesian coordinates $Y^A$ and metric $\eta_{AB}={\rm diag}(-1,-1,1,1,1,1)$.  In this same auxiliary space, $dS_{2,3}$ can be realized as the surface $\eta_{AB}Y^AY^B=+L^2$.  This manifests the common $\frak{so}(2,4)$ symmetry of the two spacetimes.  In both cases the brane breaks the symmetry to $\frak{iso}(1,3)$, so we have two effective theories that share the exact same symmetry breaking pattern yet appear to be physically distinct, with different 4-pt. amplitudes, one with a correct sign and the other a wrong sign.  Is this a counterexample to the expectation that the symmetry breaking pattern determines the EFT in the case where there is spacetime symmetry breaking?

In what follows, we will study this apparent counterexample more carefully.  The punchline is that higher-order terms in the effective theory are important in this case; the ${\cal O}\left(p^4\right)$ part of the 4-point tree amplitude gets contributions from higher curvature and extrinsic curvature terms on the brane.  When these terms are all accounted for, amplitudes in the two theories can be seen to match.  This means that the $AdS_{1,4}$ and $dS_{2,3}$ theories are actually equivalent, i.e. there exists in this case a complicated field re-definition that changes a ``wrong sign'' DBI theory into a ``correct'' sign theory.  We will find this field redefinition explicitly through the coset construction.  It also means that, at least in this example, the on-brane physics of a fixed bulk spacetime with two time dimensions is indistinguishable from that of a standard bulk spacetime, and so there should be no pathology associated with the extra time dimension.    This is not the case in the flat bulk example; in this case the higher order terms cannot contribute to the ${\cal O}\left(p^4\right)$ part of the 4-point amplitude and so the ``correct sign'' and ``wrong sign'' theories are physically distinct, as expected from their distinct symmetry breaking patterns.

\textbf{Conventions:}   We use the mostly plus metric signature and  the curvature conventions of \cite{Carroll:2004st}.  For amplitudes, we use all incoming momenta, and the Mandelstam invariants for the massless 4 particle amplitudes are defined by $s=-{1\over 2}p_1\cdot p_2$, $t=-{1\over 2}p_1\cdot p_3$, $u=-{1\over 2}p_1\cdot p_4$.  
Given a $D\times D$ matrix with components $M^\mu_{\ \nu}$, traces of matrix products are denoted using square brackets, e.g. $[M]=M^\mu_{\ \mu}$, $[M^2]=M^\mu_{\ \nu}M^\nu_{\ \mu}$, $[M^3]=M^\mu_{\ \nu}M^\nu_{\ \rho}M^\rho_{\ \mu}$, etc.
The elementary symmetric polynomials are defined as 
\be S_n[M]=M^{ [\mu_1}_{\ \mu_1}\cdots M^{ \mu_n]}_{\ \mu_n}\,,\label{symmpoldefeee}\ee
 where the anti-symmetrization is with weight 1.  We have $S_n[M]=0$ for $n>D$, and we also define $S_0[M]=1$.  
The first few are
\bea S_0(M)&=&1 \, ,\nn\\
S_1(M)&=&[M] \, ,\nn\\
S_2(M)&=&{1\over 2!}\left( [M]^2-[M^2]\right)\, ,\nn\\
S_3(M)&=&{1\over 3!}\left( [M]^3-3[M][M^2]+2[M^3]\right) \, ,\nn\\
S_4(M)&=&{1\over 4!}\left( [M]^4-6[M]^2[M^2]+8[M][M^3]+3[M^2]^2-6[M^4]\right)\, . 
\eea
Useful formulae are
\be \det(M)=S_D(M),\ \ \  \det\left(1+ M\right)=\sum_{n=0}^D S_n(M).\ee

\section{Flat brane in an AdS$_{1,4}$ bulk}

We begin by describing the EFT of a flat Minkowski 3-brane with the standard signature $(-,+,+,+)$, which we denote M$_{1,3}$, fluctuating within a fixed five dimensional bulk AdS space, with the standard signature $(-,+,+,+,+)$, which we denote AdS$_{1,4}$.  This is a standard brane-world setup in which the extra dimension is spacelike.

AdS$_{1,4}$ of radius $L$ can be realized as (the universal cover of) the surface of the hyperbola $\eta_{AB} Y^AY^B = -L^2$
 embedded into an auxiliary ambient six-dimensional two-time Minkowski space with coordinates $Y^A$, $A=0,\ldots,5$ and metric $\eta_{AB}= {\rm diag}(-1, -1, 1, 1, 1, 1)$. 
We will use Poincare Coordinates $(\rho, x^\mu)$ on AdS$_{1,4}$, $\mu=0,\ldots,3$, given by
\bea
    &&Y^0=L\cosh(\rho/L)+\frac{1}{2L} e^{\rho/L}x^2 \ ,\nn\\     
    &&Y^1= e^{\rho /L} x^0 \ ,\nn\\
    &&Y^2= L\sinh(\rho/L)-\frac{1}{2L}e^{\rho/L}x^2 \ ,\nn\\
    &&Y^{i+2}=e^{\rho /L} x^i , \ \ \  \ \ i =1,2,3 \ \ \ \,.\
\eea
Here in the expressions for $Y^0$, $Y^2$, and below, $x^2\equiv\eta_{\mu\nu}x^\mu x^\nu$ (not to be confused with the second component of $x^\mu$) where $\eta_{\mu\nu}= {\rm diag}(-1,1,1,1)$ is the flat Minkowski metric.  The AdS$_{1,4}$ metric in these coordinates reads
\be ds^2=d\rho^2+e^{2\rho/L} \eta_{\mu\nu} dx^\mu dx^\nu\,.  \ee

Following the procedure discussed in \cite{Goon:2011qf}, we take a unitary gauge in which the $x^\mu$ are used as coordinates on the brane, and so we embed the brane into AdS$_{1,4}$ via $\rho(x)=L\phi(x)$, $x^\mu(x)=x^\mu$.  The field $\phi(x)$ will be the dynamical brane bending mode.  The induced metric, inverse metric, covariant measure and extrinsic curvature are then expressed in terms of $\phi$ as follows,
\begin{equation}
   g_{\mu\nu}= e^{2\phi}\eta_{\mu\nu}+L^2 \partial_\mu\phi\partial_\nu\phi,\ \ \ g^{\mu\nu}=e^{-2\phi}\left( \eta^{\mu\nu}-{L^2 e^{-2\phi}  \gamma^2} \partial^\mu\phi\partial^\nu\phi\right) ,\ \ \sqrt{-g}={e^{4\phi}\over \gamma}, \label{Adsbraneinmeetree}
\end{equation}
\be K_{\mu\nu} ={\gamma\over L}\left(e^{2\phi}\eta_{\mu\nu}-L^2\partial_\mu\partial_\nu\phi+2L^2\partial_\mu\phi\partial_\nu\phi\right) \,, \label{braneexcuree}\ee
where 
\be \gamma\equiv {1\over \sqrt{1+L^2 e^{-2\phi}(\partial\phi)^2}}.\label{adsgammeae}\ee

The isometries of the AdS$_{1,4}$ space are induced via the 15 Lorentz generators of the six dimensional ambient space, and form the algebra $\frak{so}(2,4)$.   
Via the procedure discussed in \cite{Goon:2011qf}, they induce the following transformations on $\phi$,
\bea && P_\mu \, \phi = -\partial_\mu \phi\,  ,\ \ \ \ \nn \\ 
&&J_{\mu\nu} \, \phi = (x_\mu\partial_\nu-x_\nu\partial_\mu)\phi  \,, \nn \\
&& D\phi= -1-x^\mu\partial_\mu\phi  \, ,\nn \\
&&K_\mu\, \phi =-2x_\mu+\left[-2x_\mu x^\nu\partial_\nu+\left( x^2+L^2e^{-2\phi}\right) \partial_\mu \right]\phi  \,. \label{symmsaads}
\eea
These are the symmetry transformations of the EFT.  They satisfy the $\frak{so}(2,4)$ commutation relations
\bea
&&[D,P_\mu ]=-P_\mu \nn,\ \ \  [D , K_\mu]=K_\mu,\ \ \  [K_\mu , P_\nu]=2 J_{\mu\nu}-2\eta_{\mu\nu}D \, ,\nn \\ 
&&[J_{\mu\nu}, K_\sigma]= \eta_{\mu\sigma}K_\nu-\eta_{\nu\sigma}K_\mu ,\ \ \ [J_{\mu\nu}, P_\sigma]= \eta_{\mu\sigma}P_\nu- \eta_{\nu\sigma}P_\mu\,,   \nn\\
&& [J_{\mu\nu}, J_{\rho\sigma}]=\eta_{\mu\rho}J_{\nu\sigma}-\eta_{\nu\rho}J_{\mu\sigma}+\eta_{\nu\sigma}J_{\mu\rho}-\eta_{\mu\sigma}J_{\nu\rho}\, ,\label{so24commrelate}
\eea
with all those not shown vanishing.  
The $P_\mu$ and $J_{\mu\nu}$ are linearly realized on $\phi$ and form a subalgebra which is the four dimensional Poincare algebra $\frak{iso}(1,3)$, whereas the rest are non-linearly realized.  The symmetry breaking pattern is thus 
\be \frak{so}(2,4)\rightarrow \frak{iso}(1,3)\,. \label{adssymbpaee}\ee

Any Lagrangian constructed from the induced metric and measure \eqref{Adsbraneinmeetree}, its curvature $R_{\mu\nu\rho\sigma}$, and the extrinsic curvature \eqref{braneexcuree} will have the symmetries \eqref{symmsaads}, but as pointed out in~\cite{deRham:2010eu} there are a finite number of actions that give field equations for $\phi$ that are second order in derivatives and hence free of extra ghostly \cite{Ostrogradsky:1850fid,deUrries:1998obu} degrees of freedom.  These are known as DBI galileon terms, and they consist of all the Lovelock terms~\cite{Lovelock:1971yv} on the brane and the boundary terms corresponding to Lovelock terms in the bulk.  For our 3-brane there are five such terms\footnote{We follow here the normalizations of eq. (29) of \cite{Goon:2011qf}, with $f(\phi)=e^{\phi/L}$ followed by $\phi\rightarrow L\phi$, and with the addition of overall powers of $L$ so that the Lagrangians have mass dimension 4.}, 
\bea   
{\cal L}_1^{\rm (AdS)} &=&{1\over 4L^4} e^{4\phi} \, ,\nn\\
{\cal L}_2^{\rm (AdS)} &=&-{1\over L^4}\sqrt{- g} \ ,\nn\\
{\cal L}_3^{\rm (AdS)} &=&{1\over L^3}  \sqrt{- g}K= {1\over L^3}  \sqrt{- g} S_1(K)  \ ,\nn\\
{\cal L}_4^{\rm (AdS)} &=&-{1\over L^2} \sqrt{- g} R={1\over L^2} \sqrt{- g} \left[ -2 S_2(K) +{12\over L^2} \right]  \ ,\nn\\
{\cal L}_5^{\rm (AdS)} &=&{3\over 2 L} \sqrt{- g}\left[-{1\over 3} K^3+K_{\mu\nu}^2K-{2\over 3}K_{\mu\nu}^3-2\left( R_{\mu\nu}-\half  R  g_{\mu\nu}\right)K^{\mu\nu}\right] \nn \\
&=&{1\over L}  \sqrt{- g} \left[ 6 S_3(K) -{9\over L^2}S_1(K) \right]   \,.\ 
\label{ghostfreegentermsads} 
\eea
All of these Lagrangians except ${\cal L}_1$ are expressed using only the metric and measure \eqref{Adsbraneinmeetree}, its curvature, and the extrinsic curvature \eqref{braneexcuree}.  
In the second equalities we have made use of the Gauss-Codazzi equations,
\be R_{\mu\nu\rho\sigma}=K_{\mu\rho}K_{\nu\sigma}- K_{\mu\sigma}K_{\nu\rho} -{1\over L^2}\left( g_{\mu\rho}g_{\nu\sigma}- g_{\mu\sigma}g_{\nu\rho}  \right) \,,\label{gaussdozadse}\ee
to eliminate intrinsic curvatures in terms of extrinsic curvatures; once this is done these Lagrangians take the form of symmetric polynomials \eqref{symmpoldefeee} of the extrinsic curvature \eqref{braneexcuree}.
The term ${\cal L}_1$ is the only one that cannot written in terms of invariants; it is a Wess-Zumino term \cite{Goon:2012dy}, and can be interpreted as the 5-volume enclosed between some arbitrary reference surface and the brane \cite{Goon:2011qf}.

The terms ${\cal L}_2$ and ${\cal L}_4$ are the cosmological constant and Einstein-Hilbert terms on the brane (the only non-trivial Lovelock invariants in four dimensions).
The term ${\cal L}_3$ is the Gibbons-Hawking boundary term for a bulk Einstein-Hilbert term \cite{Gibbons:1976ue,York:1972sj,Dyer:2008hb}, and the term ${\cal L}_5$ is the boundary term for a bulk Gauss-Bonnet term. 
The Gauss-Bonnet term on the brane is a total derivative, and so we have the following total derivative combination,
\bea  {\cal L}_{\rm total\ derivative}^{\rm (AdS)}   &=& \sqrt{- g} \left( R_{\mu\nu\rho\sigma}^2- 4R_{\mu\nu}^2+R^2\right)=8 \sqrt{- g} \left[ 3 S_4(K) -{1\over L^2}S_2(K)+{3\over L^4} \right]\,. \nn\\  \label{totaldadstermeK} \eea
This can be used to eliminate the fourth order symmetric polynomial $S_4(K)$ in terms of the lower order symmetric polynomials.  Since $S_n(K)=0$ for $n>4$, this means that the non-Wess-Zumino galileons consist of precisely the non-trivial symmetric polynomials.

 In an EFT there is no a priori reason to restrict to terms such as the galileons that generate second order equations of motion, and in general all terms compatible with the symmetries should be present and arranged in a derivative expansion.  However, as discussed in \cite{Creminelli:2013fxa}, we will be concerned with equivalences between theories that ultimately derive from field redefinitions induced from some invertible change of coordinates on the coset space of the Goldstones; an invertible field redefinition should not generate new degrees of freedom and so any equivalences should be visible within the class of galileon theories.  (There are possible loopholes to this expectation, for example multi-field theories with higher-order equations that nevertheless propagate no extra degrees of freedom \cite{Gabadadze:2012tr,Motohashi:2016ftl,Klein:2016aiq,Motohashi:2018pxg}, such as the ``beyond Horndeski'' theories \cite{Zumalacarregui:2013pma,Kobayashi:2019hrl}, but this does not occur in our case and we can stay within the class of second-order theories.)

\subsection{Amplitudes\label{AdSampsection}}

The $S$ matrix is invariant under general perturbative field redefinitions, so given two theories, the matching of any given $S$-matrix element is a necessary condition for the theories to be equivalent.  We thus turn to computing some amplitudes for scattering $\phi$ fluctuations in the above AdS galileon theory, in order to compare later with the dS version of the theory.

The general galileon theory is a linear combination of the five terms \eqref{ghostfreegentermsads}
\be {\cal L}^{(\rm AdS)}=c_1 {\cal L}_1^{(\rm AdS)}+c_2 {\cal L}_2^{(\rm AdS)}+c_3 {\cal L}_3^{(\rm AdS)}+c_4 {\cal L}_4^{(\rm AdS)}+c_5 {\cal L}_5^{(\rm AdS)}\,,\label{genlagexdeadse}\ee
where $c_i$, $i={1,\ldots,5}$ are dimensionless coefficients. 

We compute amplitudes by expanding in powers of $\phi$.  Demanding absence of a linear (tadpole) term so that $\phi=0$ is a background solution imposes the following condition among the coefficients
\be c_1 - 4 c_2 + 16 c_3 - 48 c_5=0 \,. \label{adstadpolee1}\ee
Expanding to second order in the fields, we find that the mass term vanishes by \eqref{adstadpolee1}, and there is a kinetic term $-{1\over 2}{Z\over L^2}(\partial\phi)^2$ with
\be Z = c_2 - 6 c_3 + 12 c_4 - 6 c_5 \,.  \label{adscanonirmee1} \ee
In order to have a stable vacuum around which to define the perturbative $S$-matrix, we must demand $Z>0$.

At higher orders, the potential terms all vanish by \eqref{adstadpolee1} and the Lagrangian has only derivative interactions (which do not depend on $c_1$).
At cubic order, we can remove all the interactions by making the field redefinition
\be \phi\rightarrow  \phi-{1\over 2}\phi^2+{L^2\over 2Z}\left(c_3-6 c_4+9 c_5\right)(\partial\phi)^2\,. \ee
Once all this is done, the interactions start at quartic order,  
\bea {\cal L}^{\rm (AdS)}=& &-{1\over 2}{Z\over L^2} (\partial\phi)^2 +\frac{Z}{8} \left[1+{2\over Z} \left( c_3-6 c_4+9 c_5\right)\right](\partial\phi)^4 \nn\\ 
&&+\frac{1}{2}  \left[ c_4-4 c_5-\frac{1}{2 Z}\left(c_3-6 c_4+9 c_5\right){}^2\right] L^2 (\partial\phi)^2 (\partial_\mu\partial_\nu\phi)^2 \nn\\ 
&&+{\rm\ on\ shell\ trivial}+ {\cal O}\left(\phi^5\right)\,. \label{expanededformexe}\eea
Here we have used \eqref{adscanonirmee1} to elimiate $c_2$ in favor of $Z$.

We can now compute amplitudes by canonically normalizing the field, $\phi={L\over \sqrt{Z}}\hat \phi$, and using the Feynman rules. Since we have removed the cubic vertices, the 4-pt. amplitude is easy to compute: there are no exchange diagrams, so the on-shell four point amplitude comes only from a contact diagram with the vertex drawn from the $\phi^4$ terms.  There are two structures, a $p^4$ part and a $p^6$ part, coming from the four and six derivative $\phi^4$ terms respectively,
\bea {\cal A}_4^{\rm (AdS)}=&& \frac{1}{4Z} \left[1+ {2\over Z}\left(  c_3-6 c_4+9 c_5\right)\right] L^4 (s^2 + t^2 + u^2)  \nn\\ 
&& +{3\over 2 Z^2} \left[ c_4-4 c_5-\frac{1}{2 Z}\left(c_3-6 c_4+9 c_5\right){}^2\right] L^6 s t u  \,. 
\eea
We see here that the higher galileons contribute to the four point amplitude at order $p^4$.  This is unlike the case of the flat bulk DBI theory, where only ${\cal L}_2$ contributes, as mentioned in the introduction.  

The 5-pt. point amplitude is also simple to compute; since there is no factorization channel for a five point process with no on-shell non-trivial cubic interactions, the amplitude comes only from the five point contact diagram with the vertex drawn from the $\phi^5$ terms.  It contains $p^4$, $p^6$ and $p^8$ parts,  
\bea
{\cal A}_5^{\rm (AdS)} &=&  {A}_1 L^5 \sum_{1 \leq i< j}^{5} (p_i \cdot p_j)^2+ {A}_2 L^7 \sum_{1 \leq i< j}^{5} (p_i \cdot p_j)^3 \nn\\
&+& {A}_3 L^9\left[\sum_{1\leq i< j}^5(p_i \cdot p_j)^4-2 ((p_1 \cdot p_2)^2(p_3 \cdot p_4)^2+ \text{perms})\right] \,,
\eea
where
\bea
{A}_1 &=& -{2\over Z^{3/2}}\left[1+{2\over Z}\left( c_3-6 c_4+9 c_5\right)\right]  \, , \nn\\
{A}_2 &=& {12\over Z^{5/2}}\left[c_4-4c_5-{1\over 2Z}\left( c_3-6 c_4+9 c_5\right)^2\right]   \, , \nn\\
{A}_3 &=&  -{6\over Z^{5/2}}\left[ c_5 -\frac{\left(c_4-4 c_5\right) \left(c_3-6 c_4+9 c_5\right)}{Z}+ \frac{\left(c_3-6 c_4+9 c_5\right)^3}{3 Z^2}\right] \,.
\eea

\section{Flat brane in a dS$_{2,3}$
bulk }

We now consider a flat 3-brane probing a five dimensional de Sitter space with two time dimensions, which we call dS$_{2,3}$.  dS$_{2,3}$ is the maximally symmetric space with positive curvature and signature $(-,-,+++)$.  It can be realized as the surface $\eta_{AB} Y^AY^B = L^2$
 embedded into the same auxiliary ambient six-dimensional two-time Minkowski space with coordinates $Y^A$, $A=0,\ldots,5$ and metric $\eta_{AB}= {\rm diag}(-1, -1, 1, 1, 1, 1)$ that we used for AdS$_{1,4}$.
The analog of Poincare Coordinates $(\rho, x^\mu)$ on dS$_{2,3}$ are given by 
\bea
    &&Y^0=L\sinh(\rho/L)+\frac{1}{2L} e^{\rho/L}x^2 \ ,\nn\\     
    &&Y^1= e^{\rho /L} x^0 \ ,\nn\\
    &&Y^2= L\cosh(\rho/L)-\frac{1}{2L}e^{\rho/L}x^2 \ ,\nn\\
    &&Y^{i+2}=e^{\rho /L} x^i\, ,  \ \ \ \  i=1,2,3 \ \  \ \ \ \,,\
\eea
where as before, in the expressions for $Y^0$, $Y^2$ and below, $x^2\equiv\eta_{\mu\nu}x^\mu x^\nu$ with $\eta_{\mu\nu}= {\rm diag}(-1,1,1,1)$ the Minkowski 4-metric.  The dS$_{2,3}$ metric in these coordinates reads
\be ds^2=-d\rho^2+e^{2\rho/L} \eta_{\mu\nu} dx^\mu dx^\nu\,.  \ee

Embedding the brane via $\rho(x)=L \phi(x)$, $x^\mu(x)=x^\mu$, the induced metric, inverse metric, invariant measure and extrinsic curvature are expressed as follows,
\begin{equation}
   g_{\mu\nu}= e^{2\phi}\eta_{\mu\nu}-L^2 \partial_\mu\phi\partial_\nu\phi,\ \ \ g^{\mu\nu}=e^{-2\phi}\left( \eta^{\mu\nu}+{L^2 e^{-2\phi}  \gamma^2} \partial^\mu\phi\partial^\nu\phi\right) ,\ \ \sqrt{-g}={e^{4\phi}\over \gamma}, \label{branemetdsee}
\end{equation}
\be K_{\mu\nu} ={\gamma\over L}\left(e^{2\phi}\eta_{\mu\nu}+L^2\partial_\mu\partial_\nu\phi-2L^2\partial_\mu\phi\partial_\nu\phi\right) \,,\label{dspeqecurve}\ee
where 
\be \gamma\equiv {1\over \sqrt{1-L^2 e^{-2\phi}(\partial\phi)^2}}.\ee
Note that compared with the AdS case there is now a minus sign under the square roots.

The isometries of the dS$_{2,3}$ space are induced via the 15 Lorentz generators of the six dimensional ambient space, exactly the same as those which induce the isometries of AdS$_{1,4}$, and so the isometry algebra of dS$_{2,3}$ is also $\frak{so}(2,4)$.  The induced transformations on $\phi$ take the same form as \eqref{symmsaads} except for the special conformal generators $K_\mu$, which now have an opposite sign in front of the exponential term,
\bea && P_\mu \, \phi = -\partial_\mu \phi  ,\ \ \ \ \nn \\ 
&&J_{\mu\nu} \, \phi = (x_\mu\partial_\nu-x_\nu\partial_\mu)\phi  \,, \nn \\
&& D\phi= -1-x^\mu\partial_\mu\phi  ,\nn \\
&&K_\mu\, \phi =-2x_\mu+\left(-2x_\mu x^\nu\partial_\nu+\left( x^2-L^2e^{-2\phi}\right) \partial_\mu \right)\phi  \,. \label{symmsdse}
\eea
They satisfy the same $\frak{so}(2,4)$ commutation relations \eqref{so24commrelate}.  The $P_\mu$ and $J_{\mu\nu}$ are linearly realized while the rest are non-linearly realized, thus the symmetry breaking pattern is $\frak{so}(2,4)\rightarrow \frak{iso}(1,3)$, the same as for the AdS theory \eqref{adssymbpaee}.

The galileon Lagrangians are 
\bea   
{\cal L}_1^{\rm (dS)} &=&{1\over 4L^4} e^{4\phi} \, ,\nn\\
{\cal L}_2^{\rm (dS)} &=&-{1\over L^4}\sqrt{- g} \ ,\nn\\
{\cal L}_3^{\rm (dS)} &=&{1\over L^3}  \sqrt{- g}K= {1\over L^3}  \sqrt{- g} S_1(K)  \ ,\nn\\
{\cal L}_4^{\rm (dS)} &=&-{1\over L^2} \sqrt{- g} R={1\over L^2} \sqrt{- g} \left[ 2 S_2(K) -{12\over L^2} \right]  \ ,\nn\\
{\cal L}_5^{\rm (dS)} &=&{3\over 2 L} \sqrt{- g}\left[-{1\over 3} K^3+K_{\mu\nu}^2K-{2\over 3}K_{\mu\nu}^3+2\left( R_{\mu\nu}-\half  R  g_{\mu\nu}\right)K^{\mu\nu}\right] \nn \\
&=&{1\over L}  \sqrt{- g} \left[ 6 S_3(K) -{9\over L^2}S_1(K) \right]   \,.\ 
\label{ghostfreegentermsdsde} 
\eea
In the second equalities we have made use of the Gauss-Codazzi equations which now have some additional minus signs due to the fact that the extra dimension is time-like and the bulk space is positively curved,
\be R_{\mu\nu\rho\sigma}=- K_{\mu\rho}K_{\nu\sigma}+ K_{\mu\sigma}K_{\nu\rho} +{1\over L^2}\left( g_{\mu\rho}g_{\nu\sigma}- g_{\mu\sigma}g_{\nu\rho}  \right) \,.\ee
Note also that there is a different sign among the terms involving the intrinsic curvature in the first expression for ${\cal L}_5$, because of the time-like extra dimension \cite{Myers:1987yn}.

The total derivative term resulting from the Gauss-Bonnet invariant on the brane now reads,
\bea   {\cal L}_{\rm total\ derivative}^{\rm (dS)} &=&  \sqrt{- g} \left( R_{\mu\nu\rho\sigma}^2- 4R_{\mu\nu}^2+R^2\right)=8 \sqrt{- g} \left[ 3\, S_4(K) -{1\over L^2}S_2(K)+{3\over L^4} \right]\,. \nn\\  \label{totaldds393e} \eea
 
\subsection{Amplitudes\label{dSampsection}}
 
We now compute the 4-pt. and 5-pt. amplitudes in the dS theory in order to compare to those of the AdS theory in section \ref{AdSampsection}.

The general galileon theory is a linear combination
\be {\cal L}^{(\rm dS)}=d_1 {\cal L}_1^{(\rm dS)}+d_2 {\cal L}_2^{(\rm dS)}+d_3 {\cal L}_3^{(\rm dS)}+d_4 {\cal L}_4^{(\rm dS)}+d_5 {\cal L}_5^{(\rm dS)}\,,\label{genlagexdedse}\ee
where $d_i$, $i={1,\ldots,5}$ are dimensionless coefficients.  

As before we expand in powers of $\phi$.  Demanding absence of a tadpole so that $\phi=0$ is a background solution imposes the constraint
\be d_1 - 4 d_2 + 16 d_3 - 48 d_5 =0\, .\label{adstadpolee}\ee
At second order in the fields the mass term vanishes by \eqref{adstadpolee}, and there is a kinetic term $-{1\over 2}{Z\over L^2}(\partial\phi)^2$ with
\be Z = -d_2 + 6 d_3 + 12 d_4 + 6 d_5  \,,  \label{adscanonirmee} \ee
which must satisfy $Z>0$ in order for the background to be stable.

The cubic terms are eliminated by making the field redefinition
\be \phi\rightarrow   \phi-{1\over 2}\phi^2+{L^2\over 2Z}\left(d_3+6 d_4+9 d_5\right)(\partial\phi)^2  \, ,\ee
and once this is done, the Lagrangian takes the form
\bea {\cal L}^{\rm (dS)}=& &-{1\over 2}{Z\over L^2} (\partial\phi)^2 +\frac{Z}{8} \left[-1+{2\over Z} \left( d_3+6 d_4+9 d_5\right)\right](\partial\phi)^4 \nn\\ 
&&+\frac{1}{2}  \left[ d_4+4 d_5-\frac{1}{2 Z}\left(d_3+6 d_4+9 d_5\right){}^2\right] L^2 (\partial\phi)^2 (\partial_\mu\partial_\nu\phi)^2 \nn\\ 
&&+{\rm\ on\ shell\ trivial}+ {\cal O}\left(\phi^5\right)\,. \label{expanededformexeds}\eea
We have used \eqref{adscanonirmee} to elimiate $d_2$ in favor of $Z$.

For the 4-pt. amplitude we obtain
\bea {\cal A}_4^{\rm (dS)}=&& \frac{1}{4Z} \left[-1+ {2\over Z}\left(  d_3+6 d_4+9 d_5\right)\right] L^4 (s^2 + t^2 + u^2)  \nn\\ 
&& +{3\over 2 Z^2} \left[ d_4+4 d_5-\frac{1}{2 Z}\left(d_3+6 d_4+9 d_5\right){}^2\right] L^6 s t u  \,,
\eea
and for the 5-pt. amplitude we obtain
\bea
{\cal A}_5^{\rm (dS)} &=&  {A}_1 L^5 \sum_{1 \leq i< j}^{5} (p_i \cdot p_j)^2+ {A}_2 L^7 \sum_{1 \leq i< j}^{5} (p_i \cdot p_j)^3 \nn\\
&+& {A}_3 L^9\left[\sum_{1\leq i< j}^5(p_i \cdot p_j)^4-2 ((p_1 \cdot p_2)^2(p_3 \cdot p_4)^2+ \text{perms})\right] \,,
\eea
where
\bea
{A}_1 &=& -{2\over Z^{3/2}}\left[-1+{2\over Z}\left( d_3+6 d_4+9 d_5\right)\right]  \, , \nn\\
{A}_2 &=& {12\over Z^{5/2}}\left[d_4+4d_5-{1\over 2Z}\left( d_3+6 d_4+9 d_5\right)^2\right]   \, , \nn\\
{A}_3 &=&  -{6\over Z^{5/2}}\left[ d_5 -\frac{\left(d_4+4 d_5\right) \left(d_3+6 d_4+9 d_5\right)}{Z}+ \frac{\left(d_3+6 d_4+9 d_5\right)^3}{3 Z^2}\right] \,.
\eea

\section{Comparison of amplitudes}

In both the AdS$_{1,4}$ and dS$_{2,3}$ cases, the Minkowski brane breaks the symmetry $\frak{so}(2,4)$ to $\frak{iso}(1,3)$, so both theories have a single scalar degree of freedom realizing exactly the same symmetry and symmetry breaking pattern.  If the degrees of freedom, symmetry and symmetry breaking pattern determine an EFT, the AdS and dS theories should be exactly the same, despite the opposite signs within the DBI square roots, and the scattering amplitudes in section \ref{AdSampsection} should match those in section \ref{dSampsection}.

In each case, we have computed the 4-pt. amplitude, which has an order $p^4$ part and an order $p^6$ part, and the 5-pt. amplitude, which has order $p^4$, $p^6$ and $p^8$ parts.  A priori, this totals 5 different terms that must match.   We find that they all match if we take the following relations relation among the coefficients,
\bea
\left(\begin{array}{c}d_1 \\ d_2 \\ d_3 \\ d_4 \\ d_5\end{array}\right)=\left(
\begin{array}{ccccc}
 1 & 0 & 0 & 0 & 0 \\
 0 & 1 & 0 & 0 & 0 \\
 \frac{1}{8} & 0 & \frac{3}{2} & 0 & -\frac{15}{2} \\
 0 & -\frac{1}{6} & 0 & 1 & 0 \\
 \frac{1}{24} & 0 & \frac{1}{6} & 0 & -\frac{3}{2} \\
\end{array}
\right)\left(\begin{array}{c}c_1 \\ c_2 \\ c_3 \\ c_4 \\ c_5\end{array}\right),\ \ \ \label{ctodeqe1}
\eea
which has the inverse
\bea
\left(\begin{array}{c}c_1 \\ c_2 \\ c_3 \\ c_4 \\ c_5\end{array}\right)=\left(
\begin{array}{ccccc}
 1 & 0 & 0 & 0 & 0 \\
 0 & 1 & 0 & 0 & 0 \\
 \frac{1}{8} & 0 & \frac{3}{2} & 0 & -\frac{15}{2} \\
 0 & \frac{1}{6} & 0 & 1 & 0 \\
 \frac{1}{24} & 0 & \frac{1}{6} & 0 & -\frac{3}{2} \\
\end{array}
\right)\left(\begin{array}{c}d_1 \\ d_2 \\ d_3 \\ d_4 \\ d_5\end{array}\right)\,. \label{ctodeqe2}
\eea
This relation preserves the tadpole vanishing conditions \eqref{adstadpolee1}, \eqref{adstadpolee}, and also preserves the kinetic normalization constant $Z$ in \eqref{adscanonirmee1}, \eqref{adscanonirmee} once the tadpole conditions are enfored. 

With this relation, the 4-pt. and 5-pt. amplitudes in the two theories match, despite the ``wrong sign'' DBI terms in the dS$_{2,3}$ theory.  This is due to the fact that the higher-order galileons contribute to the amplitudes at lower order; for example if we include only the lowest order DBI terms ${\cal L}_{1,2}^{\rm AdS}$ in the AdS theory, as in \eqref{introadsede}, the amplitude is matched by a dS theory that has the higher order terms, whose contribution reverses the wrong sign in \eqref{introdseede}.

On closer inspection, this may be less than convincing because in each case the $p^4$ part of ${\cal A}_5$ has the same coefficient structure as the $p^4$ part of ${\cal A}_4$, and the $p^6$ part of ${\cal A}_5$ has the same coefficient structure as the $p^6$ part of ${\cal A}_4$.  As we will review in section \ref{cosetsec}, this theory is equivalent to the dilation effective theory of spontaneous conformal symmetry breaking, and these amplitude relations are a consequence of the soft theorems in this theory \cite{Boels:2015pta,Huang:2015sla,DiVecchia:2015jaq,Bianchi:2016viy,DiVecchia:2017uqn}.  Indeed, the structure of our amplitudes here matches those of the dilaton theory which can be found in \cite{Elvang:2012st,Elvang:2012yc}. We have 4 free coefficients in each theory after imposing the tadpole vanishing condition, and so it would seem that after discounting these redundant structures, there is enough freedom to equate these amplitudes, and so we should really compute more amplitudes to get a non-trivial check of equivalence.  Rather than compute more amplitudes, we will instead show directly that the two theories are equivalent by finding the invertible field re-definition that relates them.

\section{Equivalence through the coset construction\label{cosetsec}}

The coset construction \cite{Coleman:1969sm,Callan:1969sn,Volkov:1973vd} is an algorithmic method for constructing Lagrangians realizing a given symmetry breaking pattern (we refer to section 2 of \cite{Goon:2012dy} for a review of the method in our conventions).
In this section, we will see that the two different cases, the AdS and dS galileon theories, differ by a reparametrization of the coset, and thus are equivalent under a field redefinition.  This field redefinition generalizes the ``AdS/CFT equivalence transformation'' of \cite{Bellucci:2002ji}, which was applied to the galileons in \cite{Creminelli:2013fxa}.

\subsection{Weyl theory}

We start by implementing the coset construction directly on the standard $\frak{so}(2,4)$ commutation relations \eqref{so24commrelate}.  This leads to the Weyl parametrization of the theory, also known as the dilaton effective action, since it describes the dilaton of spontaneous conformal symmetry breaking.

We parametrize the coset via
\be V=e^{y\cdot P}e^{\pi D}e^{\xi\cdot  K}\,,\label{cosetweyle}\ee
where $y^\mu$ are the spacetime coordinates and $\pi, \xi^\mu$ parametrize the broken symmetry generators.
The Maurer–Cartan form is 
\be \omega= V^{-1}dV=\omega_P^\alpha P_\alpha +\omega_D D+\omega_K^\alpha K_\alpha+{1\over 2}\omega_J^{\alpha\beta} J_{\alpha\beta}\, , \label{MCformweyle}\ee
and its components are \cite{Bellucci:2002ji,McArthur:2010zm,Goon:2012dy,Hinterbichler:2012mv}
 \begin{align}
\nonumber
 \omega_{P}^{\alpha}&=e^{\pi}\rd y^\alpha\, ,\\
\nonumber
 \omega_{D}&=\rd\pi+2e^{\pi}\xi_{\mu}\rd y^{\mu}\, ,\\
\nonumber
 \omega^{\alpha}_{K}&=\rd \xi^{\alpha}+\xi^{\alpha}\rd \pi+e^{\pi}\left (2\xi^{\alpha}\xi_{\nu}\rd y^{\nu}-\xi^{2}\rd y^{\alpha}\right )\, ,\\
 \omega^{\alpha\beta}_{J}&=-4e^\pi \xi^{[\alpha}\rd y^{\beta]} \ , \label{wseylmcformexie}
 \end{align}
where the indices have been raised and lowered with the flat metric $\eta_{\mu\nu}$. 

Due to the  $\left[K_\mu,P_\nu\right]$ commutation relation we may impose the inverse Higgs constraint $\omega_D = 0$, giving the relation
\be
\xi_\mu = -\frac{1}{2}e^{-\pi}\partial_\mu\pi~.
\label{conformalinversehiggs}
\ee
Plugging back into \eqref{wseylmcformexie}, we have

\be
\begin{array}{l}
\omega_P^\alpha = e^\pi {\rm d}y^\alpha\, ,\\
\omega_K^\alpha = {1\over 2} e^{-\pi}\left(\partial_\nu\pi\partial^\alpha\pi -\partial_\nu\partial^\alpha\pi-{1\over 2}(\partial\pi)^2 \delta^\alpha_\nu\right) \rd y^\nu\, ,\\
\omega_J^{\alpha\beta} = 2\partial^{[\alpha}\pi {\rm d} y^{\beta]} \, .
\end{array}
\label{confmcforms}
\ee

The invariant vielbein $e^{\ \alpha}_{\mu}$ is extracted from $\omega_P$ via $\omega_P^\alpha \equiv e^{\ \alpha}_{\mu}\rd y^\mu$, 
\be e_\mu^{\ \alpha} = e^\pi\delta_\mu^\alpha\, .\ee
(Here and below, we use $\alpha,\beta,\dots$ to denote indices that behave like Lorentz indices, and $\mu,\nu,\ldots$ to denote indices which behave like spacetime indices.)
From this we compute the invariant metric and measure via $g_{\mu\nu}=e_\mu^{\ \alpha}e_\nu^{\ \beta}\eta_{\alpha\beta}$,
\be\label{confinmet}
g_{\mu\nu} = e^{2\pi}\eta_{\mu\nu},\ \ \ g^{\mu\nu} = e^{-2\pi}\eta^{\mu\nu},\ \ \ \sqrt{-g}=e^{4\pi}\,.
\ee

The other invariant building block in addition to the metric is the covariant derivative of the Goldstones, given via $\omega_K^\alpha =\mathcal D_\mu\xi^\alpha dy^\mu$.  Converting to space-time indices,
 $\mathcal D_\mu\xi_\nu=\left(\omega_K\right)_\mu^{\ \beta} e_\nu^{\ \alpha}\eta_{\beta\alpha}$, it becomes
\be
\mathcal D_\mu\xi_\nu =\frac{1}{2}\partial_\mu\pi\partial_\nu\pi-\frac{1}{2}\partial_\mu\partial_\nu\pi-\frac{1}{4}(\partial \pi)^2\eta_{\mu\nu}~.
\label{dxia}
\ee
Equivalently, we can use the Ricci tensor of the metric \eqref{confinmet},
 \be
R_{\mu\nu}(g) = 2\partial_\mu\pi\partial_\nu\pi-2\partial_\mu\partial_\nu\pi-\square\pi\eta_{\mu\nu}-2(\partial\pi)^2\eta_{\mu\nu}=4\mathcal D_\mu\xi_\nu+2\mathcal D_\rho\xi^\rho g_{\mu\nu}~.
\ee

The galileon Lagrangians \cite{Nicolis:2008in}, i.e. the ones that have second order equations of motion, are nothing but the symmetric polynomials made from the covariant derivative $\mathcal D_\mu\xi_\nu$, contracted and integrated with the metric and measure \eqref{confinmet}.  The only exception is ${\cal L}_3$, which is a Wess-Zumino term \cite{Goon:2012dy} and cannot be built directly in terms of the invariants (the would-be term $\propto \sqrt{-g}S_2[{\cal D}\xi]$ is a total derivative in $D=4$).  The Lagrangians are\footnote{We use the normalizations of eq. (168) of \cite{Goon:2011qf}, with $\pi\rightarrow -\pi$.}
\bea {\cal L}_1^{\rm (Weyl)}= && - {1\over 4L^4}  \sqrt{-g}= - {1\over 4L^4}  e^{4\pi}  \, ,   \nn\\
 {\cal L}_2^{\rm (Weyl)}= &&  -{1\over 12L^2 } \sqrt{-g}R=-{1 \over L^2}   \sqrt{-g} S_1[{\cal D}\xi]= - {1\over 2L^2}  e^{2\pi}(\partial\pi)^2 \,  ,   \nn\\
  {\cal L}_3^{\rm (Weyl)}= &&{1\over 2}  (\partial \pi)^2\square\pi +{1\over 4}(\partial\pi)^4 \,, \nn\\
   {\cal L}_4^{\rm (Weyl)}= && L^2   \sqrt{-g}\, {1\over 8}\left( {7\over 36} [R]^3-[R][R^2]+[R^3] \right) \nn\\ 
   =&& L^2   \sqrt{-g}\,24\, S_3[{\cal D}\xi]     \nn\\ 
   =&&  {L^2\over 2} e^{-2\pi} (\partial\pi)^2\left[ -2S_2[\partial\partial\pi]+{1\over 2}  (\partial\pi)^2\square\pi -{1\over 2} (\partial\pi)^4\right] \, ,\nn\\
   {\cal L}_5^{\rm (Weyl)}= && L^4   \sqrt{-g}\, {1\over 192}\left(  {31\over 18}[R]^4-13 [R^2][R]^2+9[R^2]^2+20[R][R^3]-18[R^4]  \right) \nn\\ 
    = && L^4   \sqrt{-g}\,96\, S_4[{\cal D}\xi]     \nn\\
    =&& {L^4\over 2} e^{-4\pi} (\partial\pi)^2\left[ 6S_3[\partial\partial\pi]-6 (\partial\pi)^2S_2[\partial\partial\pi]  +5 (\partial\pi)^4 \square\pi -{14\over 4} (\partial\pi)^6\right] \,.\nn\\ \label{weyllagsere}
\eea
We have normalized the Lagrangians with the scale $L$ so as to aid the comparison to the AdS and dS theories, but this is not a scale intrinsic to the construction at this point.

The symmetry transformations induced by the transformations on the coset are
\bea && P_\mu \, \pi = -\partial_\mu \pi  ,\ \ \ \ \\ 
&&J_{\mu\nu} \, \pi = (x_\mu\partial_\nu-x_\nu\partial_\mu)\pi  \,, \\
&& D\pi= -1-x^\mu\partial_\mu\pi  ,\\
&&K_\mu\, \pi =-2x_\mu+\left(-2x_\mu x^\nu\partial_\nu+x^2 \partial_\mu \right)\pi \,.
\eea
These are symmetries of the Lagrangians \eqref{weyllagsere}.  They satisfy the commutations relations \eqref{so24commrelate}.

Note that, unlike the DBI theories, these Lagrangians \eqref{weyllagsere} appear as a pure derivative expansion, i.e. the $n$-th Lagrangian contains only terms with $2n-2$ derivatives.  This means that the power counting for figuring out which terms contribute to the amplitudes at any given order in the external momenta is simple.  For example, the order $p^4$ part of the 4-pt. amplitude famously \cite{Komargodski:2011vj} only gets contributions from the 4 derivative Wess-Zumino term $ {\cal L}_3^{\rm (Weyl)}$.   In the DBI theories on the other hand, the power counting gets mixed up between the terms and, as we have seen, the same order $p^4$ 4-pt. amplitude gets contributions from the higher order terms.

\subsection{AdS theory}

As shown in \cite{Bellucci:2002ji}, the AdS DBI theory is equivalent to the Weyl theory, and is obtained by choosing a different parametrization of the coset \eqref{cosetweyle}.  It was later shown that the galileon terms are mapped into each other under this reparametrization \cite{Creminelli:2013fxa}.   Here we will review how this works, and we will use a slightly different parametrization of the coset that makes the smooth limit between the two more transparent and shows how to extend to the dS case.  

To get the AdS theory we change the basis in the algebra from $\{ K_\mu,D,P_\mu,J_{\mu\nu}\} $ to $\{ \hat K_\mu,D,P_\mu,J_{\mu\nu}\}$, where
\be \hat K^\mu= K^\mu+L^2 P^\mu\,.\label{Adstheparece}\ee
In this basis the commutators \eqref{so24commrelate} become
\bea
&&[ D,P_\mu ]=- P_\mu \nn,\ \ \  [ D , \hat K_\mu]=\hat K_\mu -{2L^2 } P_\mu,\ \ \  [\hat K_\mu , P_\nu]=2 J_{\mu\nu}-2 \eta_{\mu\nu} D\,,  \ \  [\hat K_\mu , \hat K_\nu]= 4L^2 J_{\mu\nu}\,, \nn \\ 
&&[J_{\mu\nu}, \hat K_\sigma]= \eta_{\mu\sigma}\hat K_\nu-\eta_{\nu\sigma}\hat K_\mu ,\ \ \ [J_{\mu\nu}, P_\sigma]= \eta_{\mu\sigma}P_\nu- \eta_{\nu\sigma}P_\mu \,, \nn\\
&& [J_{\mu\nu}, J_{\rho\sigma}]=\eta_{\mu\rho}J_{\nu\sigma}-\eta_{\nu\rho}J_{\mu\sigma}+\eta_{\nu\sigma}J_{\mu\rho}-\eta_{\mu\sigma}J_{\nu\rho}\,,
\eea
with all others vanishing.  The advantage of this parametrization over that of \cite{Bellucci:2002ji,Creminelli:2013fxa} will be that we can transparently see how the original basis, and hence the Weyl theory, is recovered as $L^2\rightarrow 0$, and how the dS theory is obtained from $L^2\rightarrow -L^2$.

We now parametrize the coset as
\be V=e^{x\cdot P}e^{\phi  D}e^{\Lambda\cdot \hat K}\,,\ee
where $x^\mu$ are the spacetime coordinates and $\phi, \Lambda^\mu$ parametrize the broken directions.

The Maurer–Cartan form is 
\be \omega= V^{-1}dV=\hat\omega_P^\alpha P_\alpha +\hat\omega_{ D} D+\hat\omega_{\hat K}^\alpha \hat K_\alpha+{1\over 2}\hat\omega_J^{\alpha\beta} J_{\alpha\beta}\,, \label{MCformadsee} \ee
and its components are 
\bea
&&\hat\omega_{ D}= {1-L^2\lambda^2\over 1+L^2\lambda^2}\left( d\phi+{2 } e^{\phi}   {\lambda_\mu \over 1-L^2\lambda^2}dx^\mu\right)\, ,  \\
&&\hat\omega_P^\alpha= -{2L^2\over 1+L^2 \lambda^2}\lambda^\alpha d\phi+ e^{\phi}\left( dx^\alpha  -{2L^2\over 1+L^2\lambda^2}  \lambda^\alpha \lambda_\mu dx^\mu \right) \, ,  \\
&&\hat\omega_{\hat K}^\alpha= {1\over 1+L^2 \lambda^2}\left[ d\lambda^\alpha+\lambda^\alpha d\phi+ e^{\phi}\left( -\lambda^2 dx^\alpha +2 \lambda^\alpha\lambda_\mu dx^\mu\right)\right] \, , \\
&&\hat\omega_J^{\alpha\beta} =-{4\over 1+L^2 \lambda^2}\left(  L^2\lambda^{[\alpha}d\lambda^{\beta]}+e^{\phi}  \lambda^{[\alpha}dx^{\beta]}  \right) \, ,
\eea
where
\be \lambda^\mu\equiv {1\over L}\Lambda^\mu{ \tan(L \Lambda)\over \Lambda},\ \ \ \Lambda=\sqrt{\Lambda_\mu \Lambda^\mu}\,.\ee
Note that these reduce to the corresponding Weyl theory quantities in \eqref{wseylmcformexie} as $L^2\rightarrow 0$.

We impose the inverse Higgs constraint $\omega_{ D}=0$, which has the solution\footnote{There is another solution with a minus sign in front of the square root, but we choose this one because it matches to the solution \eqref{conformalinversehiggs} when $L^2\rightarrow 0$.  Choosing the other solution would put a minus sign in front of the extrinsic curvature in \eqref{dxia2}.
},
\be 	\lambda_\mu=-{ e^{-\phi}\partial_\mu \phi\over 1+ \sqrt{ 1+L^2 e^{-2\phi}(\partial \phi)^2}} \,.\ee

The invariant vielbein $e^{\ \alpha}_{\mu}$ is extracted from $\omega_P$ via $\omega_P^\alpha \equiv e^{\ \alpha}_{\mu}\rd y^\mu$, 
\be e_\mu^{\ \alpha} =e^{\phi} \delta_\mu^\alpha-{2L^2\over 1+L^2 \lambda^2}\left( e^{\phi}\lambda^\alpha  \lambda_\mu +\lambda^\alpha \partial_\mu \phi\right)  ,\ee
which gives the invariant metric via $g_{\mu\nu}=e_\mu^{\ \alpha}e_\nu^{\ \beta}\eta_{\alpha\beta}$,
\be\label{confinmet2}
g_{\mu\nu} =  e^{2\phi}\eta_{\mu\nu}+L^2\partial_\mu \phi\partial_\nu \phi\,,
\ee
which is nothing but the brane-induced metric in \eqref{Adsbraneinmeetree}.  It reduces to the Weyl metric in \eqref{confinmet} as $L^2\rightarrow 0$.

The covariant derivative of the Goldstones, the invariant building block, is given through $\mathcal D_\mu\Lambda_\nu=\left(\omega_K\right)_\mu^{\ \beta} e_\nu^{\ \alpha}\eta_{\beta\alpha}$, and takes the form
\be
\mathcal D_\mu\Lambda_\nu ={1\over 2L}\left( K_{\mu\nu}-{1\over L}g_{\mu\nu}\right)\,,
\label{dxia2}
\ee
where $K_{\mu\nu}$ is the extrinsic curvature in \eqref{braneexcuree}.   It reduces to its Weyl counterpart \eqref{dxia} as $L^2\rightarrow 0$.

The AdS Lagrangians \eqref{ghostfreegentermsads} can all be expressed in terms of symmetric polynomials of this invariant, with the exception of ${\cal L}_1^{\rm (AdS)}$ which is a Wess-Zumino term,
\bea   
{\cal L}_1^{\rm (AdS)} &=&{1\over 4 L^4} e^{4\phi} \, ,\nn\\
{\cal L}_2^{\rm (AdS)} &=&- {1\over  L^4} \sqrt{- g} \ ,\nn\\
{\cal L}_3^{\rm (AdS)} &=& {1\over L^2}  \sqrt{- g} \left[2\,S_1({\cal D}\Lambda)+{4\over L^2}\right]  \ ,\nn\\
{\cal L}_4^{\rm (AdS)} &=& \sqrt{- g} \left[ -8\,S_2({\cal D}\Lambda)-{12\over L^2}S_1({\cal D}\Lambda)\right]  \ ,\nn\\
{\cal L}_5^{\rm (AdS)} &=&L^2 \sqrt{- g} \left[ 48\, S_3({\cal D}\Lambda) +{48\over L^2} S_2({\cal D}\Lambda)+{18\over L^4} S_1({\cal D}\Lambda) -{12\over L^6} \right]   \,.\ 
\label{ghostfreegentermsadsxie} 
\eea
In addition, the total derivative \eqref{totaldadstermeK} becomes
\bea   {\cal L}_{\rm total\ derivative}^{\rm (AdS)} &=&L^4 \sqrt{- g} \left[ 384\, S_4({\cal D}\Lambda) +{192\over L^2}S_3({\cal D}\Lambda)+{64\over L^4}S_2({\cal D}\Lambda) \right]\,,  \label{totaldadstermeK2} \eea
and can be used to eliminate $S_4({\cal D}\Lambda) $ in favor of the lower order symmetric polynomials.

The symmetry transformations induced by the transformations on the coset are precisely those in \eqref{symmsaads}, and they satisfy the commutations relations \eqref{so24commrelate}.

\subsection{dS theory}

To find a coset parametrization for the dS theory, we simply take $L^2\rightarrow -L^2$ in \eqref{Adstheparece},
\be \hat K^\mu= K^\mu-L^2 P^\mu\,,\ee
which now satisfies the commutators
\bea
&&[ D,P_\mu ]=- P_\mu \nn,\ \ \  [ D , \hat K_\mu]=\hat K_\mu +{2L^2 } P_\mu,\ \ \  [\hat K_\mu , P_\nu]=2 J_{\mu\nu}-2 \eta_{\mu\nu} D\,,  \ \  [\hat K_\mu , \hat K_\nu]= -4L^2 J_{\mu\nu}\,, \nn \\ 
&&[J_{\mu\nu}, \hat K_\sigma]= \eta_{\mu\sigma}\hat K_\nu-\eta_{\nu\sigma}\hat K_\mu ,\ \ \ [J_{\mu\nu}, P_\sigma]= \eta_{\mu\sigma}P_\nu- \eta_{\nu\sigma}P_\mu\,,  \nn\\
&& [J_{\mu\nu}, J_{\rho\sigma}]=\eta_{\mu\rho}J_{\nu\sigma}-\eta_{\nu\rho}J_{\mu\sigma}+\eta_{\nu\sigma}J_{\mu\rho}-\eta_{\mu\sigma}J_{\nu\rho} \,.
\eea

We parametrize the coset as
\be V=e^{x\cdot P}e^{\phi  D}e^{\Lambda\cdot \hat K}\, ,\ee
the Maurer–Cartan form is 
\be \omega= V^{-1}dV=\hat\omega_P^\alpha P_\alpha +\hat\omega_{ D} D+\hat\omega_{\hat K}^\alpha \hat K_\alpha+{1\over 2}\hat\omega_J^{\alpha\beta} J_{\alpha\beta}\, ,\label{MCformdseee}\ee
and its components are
\bea
&&\hat \omega_{\hat D}= {1+L^2\lambda^2\over 1-L^2\lambda^2}\left( d\phi+{2} e^{\phi}   {\lambda_\mu \over 1+L^2\lambda^2}dx^\mu\right)  \,, \\
&&\hat \omega_P^\alpha= {2L^2\over 1-L^2\lambda^2}\lambda^\alpha d\phi+ e^{\phi}\left( dx^\alpha  +{2L^2\over 1-L^2\lambda^2}  \lambda^\alpha \lambda_\mu dx^\mu \right)\,,   \\
&&\hat \omega_{\hat K}^\alpha= {1\over  1-L^2\lambda^2}\left[ d\lambda^\alpha+\lambda^\alpha d\phi+ e^{\phi}\left(- \lambda^2 dx^\alpha +2 \lambda^\alpha\lambda_\mu dx^\mu\right)\right] \,, \\
&&\hat \omega_J^{\alpha\beta} ={4\over 1-L^2\lambda^2}\left( L^2 \lambda^{[\alpha}d\lambda^{\beta]}-e^{\phi}  \lambda^{[\alpha}dx^{\beta]}  \right)  \,,
\eea
where
\be \lambda^\mu\equiv {1\over L} \Lambda^\mu{ \tanh(L\Lambda)\over \Lambda},\ \ \ \Lambda=\sqrt{\Lambda_\mu \Lambda^\mu}\,.\ee

The inverse Higgs constraint is $\omega_{ D}=0$ and the solution which smoothly reproduces the corresponding Weyl theory solution in \eqref{conformalinversehiggs} as $L^2\rightarrow 0$ is
\be 	\lambda_\mu=-{e^{-\phi}\partial_\mu \phi\over 1+ \sqrt{ 1-L^2e^{-2\phi}(\partial \phi)^2}} \,.\ee

The invariant vielbein is 
\be e_\mu^{\ \alpha} =e^{\phi} \delta_\mu^\alpha+{2L^2\over 1-L^2\lambda^2}\left( e^{\phi}\lambda^\alpha  \lambda_\mu +\lambda^\alpha \partial_\mu \phi\right)  ,\ee
and the invariant metric is
\be\label{confinmet3}
g_{\mu\nu} =  e^{2\phi}\eta_{\mu\nu}-L^2\partial_\mu \phi\partial_\nu \phi\,,
\ee
which is the same as the brane-induced metric \eqref{branemetdsee}.

The covariant derivative of the Goldstones, the invariant building block, is given by
\be
\mathcal D_\mu\Lambda_\nu ={1\over 2L}\left(- K_{\mu\nu}+{1\over L}g_{\mu\nu}\right)\,,
\label{dxia3}
\ee
where $K_{\mu\nu}$ is the extrinsic curvature \eqref{dspeqecurve}.

The dS Lagrangians \eqref{ghostfreegentermsdsde} can all be expressed in terms of symmetric polynomials of this invariant, with the exception of ${\cal L}_1^{\rm (dS)}$ which is a Wess-Zumino term,
\bea   
{\cal L}_1^{\rm (dS)} &=&{1\over 4 L^4} e^{4\phi} \, ,\nn\\
{\cal L}_2^{\rm (dS)} &=&- {1\over  L^4} \sqrt{- g} \ ,\nn\\
{\cal L}_3^{\rm (dS)} &=& {1\over L^2}  \sqrt{- g} \left[-2\,S_1({\cal D}\Lambda)+{4\over L^2}\right]  \ ,\nn\\
{\cal L}_4^{\rm (dS)} &=& \sqrt{- g} \left[ 8\,S_2({\cal D}\Lambda)-{12\over L^2}S_1({\cal D}\Lambda)\right]  \ ,\nn\\
{\cal L}_5^{\rm (dS)} &=&L^2 \sqrt{- g} \left[ -48\, S_3({\cal D}\Lambda) +{48\over L^2} S_2({\cal D}\Lambda)-{18\over L^4} S_1({\cal D}\Lambda) -{12\over L^6} \right]   \,.\ 
\label{ghostfreegentermsadsxie} 
\eea

The total derivative \eqref{totaldds393e} becomes
\bea   {\cal L}_{\rm total\ derivative}^{\rm (dS)} &= &L^4  \sqrt{- g} \left[ 384 S_4({\cal D}\Lambda) -{192\over L^2}S_3({\cal D}\Lambda)+{64\over L^4}S_2({\cal D}\Lambda) \right]\,.  \label{totaldadstermeK3} \eea

The symmetry transformations induced by the transformations on the coset are precisely those in \eqref{symmsdse}, and they satisfy the commutations relations \eqref{so24commrelate}.

\subsection{Mapping Weyl to AdS}

The find the relation between the Weyl and AdS theories, we equate the Maurer–Cartan forms \eqref{MCformweyle} and \eqref{MCformadsee}.  Expressing $\hat K_\mu$ in terms of $K_\mu,P_\mu$ using \eqref{Adstheparece} and reading off the coefficients of the algebra elements $P_\alpha$, $K_\alpha$, $D$, $J_{\alpha\beta}$ respectively gives
\bea && \omega_P^\alpha= \hat\omega_P^\alpha+L^2  \hat\omega_{\hat K}^\alpha \, ,\label{coefalre}\\
&& \omega_K^\alpha= \hat\omega_{\hat K}^\alpha  \, ,\ \ \  \omega_D^\alpha= \hat\omega_{D}^\alpha  \, ,\ \ \  \omega_J^{\alpha\beta}= \hat\omega_{J}^{\alpha\beta}  \, .
\eea

One can check that these are all satisfied by the following change of coordinates on the coset \cite{Bellucci:2002ji},
\be y^\mu=x^\mu +L^2e^{-\phi}\lambda^\mu,\ \ \ e^{\pi}={e^{\phi}\over 1+L^2\lambda^2},\ \ \ \xi^\mu=\lambda^\mu,\ee
or inversely,
\be x^\mu=y^\mu-L^2{e^{-\pi}   \over  1+{ L^2}\xi^2}\xi^\mu\,,  \ \ \  e^{\phi}= \left(1+{ L^2}\xi^2\right) e^\pi, \ \ \lambda^\mu= \xi^\mu \,.\label{xtoyrelee}\ee
This is the field redefinition that connects the Weyl and AdS theory\footnote{There is a potential subtlety in that the inverse Higgs constraints for different coset parametrizations may not map into each other.  This does not affect the cases studied here, but can be an issue in other cases such as those with higher co-dimension branes \cite{Klein:2017npd}.}.  Note that the coordinates also change by a field-dependent transformation, so if expanded out the transformation would involve an infinite series in powers of the fields and derivatives.   The transformation is however explicitly invertible, it reduces to the identity as $L^2\rightarrow 0$, and it preserves the vacuum solution $\phi=\pi=0$.  Furthermore, if the fields fall to zero asymptotically, as for example in $S$-matrix scattering experiments, the coordinates $x^\mu$ and $y^\mu$ asymptotically become equal to each other exponentially fast, and so this re-definition will preserve the $S$-matrix.

We can find the transformation of the vierbein and measure from the relation \eqref{coefalre} (here we use hats to distinguish the AdS vierbein and metric from the Weyl vierbein and metric),
\be {e}_\mu^{ \ \alpha} dy^\mu= \hat e_\mu^{\ \alpha} dx^\mu +L^2 {\cal D}_\mu \Lambda^\alpha dx^\mu=\left(\delta_\mu^\nu+L^2 {\cal D}_\mu \Lambda^\nu\right)  \hat e_\nu^{\ \alpha} dx^\mu  ,\ee
\be \sqrt{-g}d^Dy= \det(e_\mu^{\ \alpha})d^Dy= \det \left(\delta_\mu^\nu+L^2 {\cal D}_\mu \Lambda^\nu\right)  \det(\hat e_\mu^{\ \alpha})d^Dx= \det \left(\delta_\mu^\nu+L^2 {\cal D}_\mu \Lambda^\nu\right)  \sqrt{-\hat g} d^Dx .\label{measuretraadnsfee}\ee

The transformation of the invariant building blocks can be found from,
\bea  && \omega_K^\alpha = \hat\omega_{\hat K}^\alpha\  \Rightarrow \ {\cal D}_\mu \xi^\alpha dy^\mu=  {\cal D}_\mu \Lambda^\alpha dx^\mu\  \Rightarrow \  {\cal D}_\beta \xi^\alpha e_\mu^{\ \beta} dy^\mu =  {\cal D}_\beta \Lambda^\alpha \hat e_\mu^{\ \beta} dx^\mu   \nn\\
&& \Rightarrow\  {\cal D}_\beta \xi^\alpha \left(\delta_\mu^\nu+L^2 {\cal D}_\mu \Lambda^\nu\right)  \hat e_\nu^{\ \beta} dx^\mu=  {\cal D}_\beta \Lambda^\alpha \hat e_\mu^{\ \beta} dx^\mu\ \Rightarrow\ {\cal D}_\beta \xi^\alpha \left(\delta_\gamma^\beta+L^2 {\cal D}_\gamma \Lambda^\beta \right)  \hat e_\mu^{\ \gamma} dx^\mu=  {\cal D}_\beta \Lambda^\alpha \hat e_\mu^{\ \beta} dx^\mu \nn\\
&& \Rightarrow  \left(\delta_\beta^\gamma+L^2 {\cal D}_\beta \Lambda^\gamma \right)  {\cal D}_\gamma \xi^\alpha   =  {\cal D}_\beta \Lambda^\alpha \,.
\eea
In matrix notation, this becomes
\be {\cal D}\xi={1\over 1+L^2 {\cal D}\Lambda} {\cal D}\Lambda,\ \ \ {\cal D}\Lambda={1\over 1-L^2 {\cal D}\xi} {\cal D}\xi, \label{invariantrelationadese}\ee
manifesting how they become equal in the limit $L^2\rightarrow 0$.
Note that we have expressed the  relation between building blocks in the form in which they have only Lorentz indices, because the relation takes its simplest form in this case.  Since the Lagrangians are always expressed as traces of contractions of these building blocks, they can be freely traded for their Lorentz indexed form.

We can now use these relations to map the Weyl and AdS galileon terms into each other, as described in \cite{Creminelli:2013fxa}.  Starting with the AdS galileons written in terms of ${\cal D}\Lambda$ \eqref{ghostfreegentermsadsxie}, we use \eqref{invariantrelationadese} to transform the invariants and \eqref{measuretraadnsfee} to transform the measure.  The result can then be written in terms of the Weyl galileons expressed in terms of ${\cal D}\xi$ \eqref{weyllagsere}.  
This does not work for ${\cal L}_1^{\rm (AdS)}$ since it cannot be expressed in terms of the invariants, so in this case we use \eqref{xtoyrelee} directly to transfrom $e^{4\phi}\rightarrow\left(1+{ L^2}\xi^2\right)^4 e^{4\pi}$, and to compute 
\be {dx^\mu\over dy^\nu }=\delta^\mu_\nu-L^2\partial_\nu\left({e^{-\pi}   \over  1+{ L^2}\xi^2}\xi^\mu\right)\, ,\ee 
and use it to transform the measure $d^4x=\det \left( dx^\mu\over dy^\nu\right)  d^4y.$  The result can be expressed as a combination of the Weyl galileons.  

The transformations can be summarized as\footnote{The numerical values in our matrices \eqref{AdStoweymdsatre} and \eqref{AdStoweymatreads2} differ from those in \cite{Creminelli:2013fxa} due to our different choices for normalizing the Lagrangians.  We agree with their values when changing to their normalizations.},
\be \left(\begin{array}{c}{\cal L}_1^{\rm(AdS)} \\ {\cal L}_2^{\rm(AdS)} \\ {\cal L}_3^{\rm(AdS)} \\ {\cal L}_4^{\rm(AdS)} \\ {\cal L}_5^{\rm(AdS)}\end{array}\right)=\left(
\begin{array}{ccccc}
 -1 & \frac{1}{2} & -\frac{1}{8} & \frac{1}{48} & -\frac{1}{384} \\
 4 & -1 & 0 & \frac{1}{24} & -\frac{1}{96} \\
 -16 & 2 & 0 & \frac{1}{12} & -\frac{1}{24} \\
 0 & 12 & 0 & -\frac{1}{2} & 0 \\
 48 & -30 & 0 & -\frac{5}{4} & \frac{1}{8} \\
\end{array}
\right) \left(\begin{array}{c}{\cal L}_1^{\rm(Weyl)} \\ {\cal L}_2^{\rm(Weyl)} \\ {\cal L}_3^{\rm(Weyl)} \\ {\cal L}_4^{\rm(Weyl)} \\ {\cal L}_5^{\rm(Weyl)}\end{array}\right) \,.  \label{AdStoweymdsatre}\ee 
Note that ${\cal L}^{\rm (AdS)}_{2,3,4,5}$ do not contribute to the Wess-Zumino term ${\cal L}^{\rm (Weyl)}_{3}$ since $S_2({\cal D}\xi)$ vanishes as a total derivative in $D=4$.  Only the AdS Wess-Zumino term ${\cal L}^{\rm (AdS)}_{1}$ contributes to ${\cal L}^{\rm (Weyl)}_{3}$, as expected since an invertible map should map non-Wess-Zumino terms only into other non-Wess-Zumino terms, and Wess-Zumino terms should be mapped into each other modulo non-Wess-Zumino terms (note that Wess-Zumino-ness is only well defined modulo non-Wess-Zumino terms.)  The mapping from Weyl to AdS galileons can be obtained by inverting \eqref{AdStoweymdsatre},
\be  \left(\begin{array}{c}{\cal L}_1^{\rm(Weyl)} \\ {\cal L}_2^{\rm(Weyl)} \\ {\cal L}_3^{\rm(Weyl)} \\ {\cal L}_4^{\rm(Weyl)} \\ {\cal L}_5^{\rm(Weyl)}\end{array}\right)=\left(
\begin{array}{ccccc}
 0 & \frac{1}{8} & -\frac{5}{128} & \frac{1}{96} & -\frac{1}{384} \\
 0 & 0 & -\frac{1}{16} & \frac{1}{24} & -\frac{1}{48} \\
 -8 & 0 & \frac{1}{8} & 0 & -\frac{1}{8} \\
 0 & 0 & -\frac{3}{2} & -1 & -\frac{1}{2} \\
 0 & -48 & -15 & -4 & -1 \\
\end{array}
\right) \left(\begin{array}{c}{\cal L}_1^{\rm (AdS)} \\ {\cal L}_2^{\rm(AdS)} \\ {\cal L}_3^{\rm(AdS)} \\ {\cal L}_4^{\rm(AdS)} \\ {\cal L}_5^{\rm(AdS)}\end{array}\right)\,.   \label{AdStoweymatreads2}\ee
Parts of the inverse can also be checked directly by transforming the non-Wess-Zumino Weyl galileons, but as pointed out in \cite{Creminelli:2013fxa}, the total derivative \eqref{totaldadstermeK2} must be used to re-express the terms quartic in ${\cal D}\Lambda$ in favor of the lower order terms.

Despite the fact that the matrices in \eqref{AdStoweymdsatre}, \eqref{AdStoweymatreads2} contain only fixed pure numbers, this mapping is actually a one parameter family of maps parametrized by $L^2$.  This is obscured by our decision to use $L$ to set the scale of the Weyl and AdS Lagrangians.  For example, suppose we choose a new independent mass scale, say $\Lambda$, to parametrize the Weyl Lagrangians, and ask for the transformation of the Weyl kinetic term ${\cal L}_2^{\rm(Weyl)}$ normalized by $\Lambda$,
\be -{1\over 2}\Lambda^2 e^{2\pi}(\partial\pi)^2=-{1\over 16} \Lambda^2L^2 {\cal L}_3^{\rm (AdS)}+{1\over 24} \Lambda^2L^2 {\cal L}_4^{\rm (AdS)}-{1\over 48} \Lambda^2L^2 {\cal L}_5^{\rm (AdS)}\,.\ee
The righthand side now has a regular limit as $L^2\rightarrow 0$ which reproduces the Weyl term, followed by an expansion in powers of $L^2$, 
\be -{1\over 2}\Lambda^2 e^{2\pi}(\partial\pi)^2=-{1\over 2}\Lambda^2 e^{2\phi}(\partial\phi)^2+{\cal O}\left(L^2\right)\,.\ee
It is clear that the map  thus gives a one parameter deformation of the Weyl theory parametrized by $L^2$, and since the transformation ultimately stems from a field redefinition, $L^2$ is a redundant parameter.  In particular, in our example with only the kinetic term ${\cal L}_2^{\rm(Weyl)}$, the right-hand side is secretly a free theory, since the left-hand side is, and so it is clear that the scale $L^2$ is a fake scale.  Since the deformation is analytic in the redundant parameter $L^2$, there is nothing preventing us from taking it to be negative, and this is how we arrive at the dS theory.

\subsection{Mapping Weyl to dS}

To find the relation between the Weyl and dS theories, we equate the Maurer–Cartan forms \eqref{MCformweyle} and \eqref{MCformdseee},
\bea && \omega_P^\alpha= \hat\omega_P^\alpha-L^2  \hat\omega_{\hat K}^\alpha \, ,\label{coefalrdse}\\
&& \omega_K^\alpha= \hat\omega_{\hat K}^\alpha  \, ,\ \ \  \omega_D^\alpha= \hat\omega_{D}^\alpha  \, ,\ \ \  \omega_J^{\alpha\beta}= \hat\omega_{J}^{\alpha\beta}  \, .
\eea
These are all satisfied by the following change of coordinates on the coset,
\be y^\mu=x^\mu -L^2e^{-\phi}\lambda^\mu,\ \ \ e^{\pi}={e^{\phi}\over 1-L^2\lambda^2},\ \ \ \xi^\mu=\lambda^\mu,\ee
or inversely,
\be x^\mu=y^\mu+L^2{e^{-\pi}   \over  1-{ L^2}\xi^2}\xi^\mu\,,  \ \ \  e^{\phi}= \left(1-{ L^2}\xi^2\right) e^\pi, \ \ \lambda^\mu= \xi^\mu \,.\label{xtoyreleeds}\ee

The transformation of the measure is (the hatted metric is the dS metric),
\be \sqrt{-g}d^Dy= \det \left(\delta_\mu^\nu-L^2 {\cal D}_\mu \Lambda^\nu\right)  \sqrt{-\hat g} d^Dx \,,\label{measuretrdsansfee}\ee
and the transformation of the invariant building block is,
\be  \left(\delta_\beta^\gamma-L^2 {\cal D}_\beta \Lambda^\gamma \right)  {\cal D}_\gamma \xi^\alpha   =  {\cal D}_\beta \Lambda^\alpha \,, \ee
or in matrix notation,
\be {\cal D}\xi={1\over 1-L^2 {\cal D}\Lambda} {\cal D}\Lambda,\ \ \ {\cal D}\Lambda={1\over 1+L^2 {\cal D}\xi} {\cal D}\xi\, . \label{invariantrelationese}\ee

The transformations from dS to Weyl galileons is
\be \left(\begin{array}{c}{\cal L}_1^{\rm(dS)} \\ {\cal L}_2^{\rm(dS)} \\ {\cal L}_3^{\rm(dS)} \\ {\cal L}_4^{\rm(dS)} \\ {\cal L}_5^{\rm(dS)}\end{array}\right)=\left(
\begin{array}{ccccc}
 -1 & -\frac{1}{2} & -\frac{1}{8} & -\frac{1}{48} & -\frac{1}{384} \\
 4 & 1 & 0 & -\frac{1}{24} & -\frac{1}{96} \\
 -16 & -2 & 0 & -\frac{1}{12} & -\frac{1}{24} \\
 0 & 12 & 0 & -\frac{1}{2} & 0 \\
 48 & 30 & 0 & \frac{5}{4} & \frac{1}{8} \\
\end{array}
\right)\left(\begin{array}{c}{\cal L}_1^{\rm(Weyl)} \\ {\cal L}_2^{\rm(Weyl)} \\ {\cal L}_3^{\rm(Weyl)} \\ {\cal L}_4^{\rm(Weyl)} \\ {\cal L}_5^{\rm(Weyl)}\end{array}\right)  \, , \label{AdStoweymatre}\ee 
and the inverse is
\be  \left(\begin{array}{c}{\cal L}_1^{\rm(Weyl)} \\ {\cal L}_2^{\rm(Weyl)} \\ {\cal L}_3^{\rm(Weyl)} \\ {\cal L}_4^{\rm(Weyl)} \\ {\cal L}_5^{\rm(Weyl)}\end{array}\right)=\left(
\begin{array}{ccccc}
 0 & \frac{1}{8} & -\frac{5}{128} & -\frac{1}{96} & -\frac{1}{384} \\
 0 & 0 & \frac{1}{16} & \frac{1}{24} & \frac{1}{48} \\
 -8 & 0 & \frac{1}{8} & 0 & -\frac{1}{8} \\
 0 & 0 & \frac{3}{2} & -1 & \frac{1}{2} \\
 0 & -48 & -15 & 4 & -1 \\
\end{array}
\right) \left(\begin{array}{c}{\cal L}_1^{\rm (dS)} \\ {\cal L}_2^{\rm(dS)} \\ {\cal L}_3^{\rm(dS)} \\ {\cal L}_4^{\rm(dS)} \\ {\cal L}_5^{\rm(dS)}\end{array}\right) \,.  \label{AdStoweymatre2}\ee

\subsection{Mapping AdS to dS}

The mapping from AdS to dS galileons can now be obtained by composing the map from AdS to Weyl and the map from Weyl to dS.  The result is

\be  \left(\begin{array}{c}{\cal L}_1^{\rm(dS)} \\ {\cal L}_2^{\rm(dS)} \\ {\cal L}_3^{\rm(dS)} \\ {\cal L}_4^{\rm(dS)} \\ {\cal L}_5^{\rm(dS)}\end{array}\right)=\left(
\begin{array}{ccccc}
 1 & 0 & \frac{1}{8} & 0 & \frac{1}{24} \\
 0 & 1 & 0 & \frac{1}{6} & 0 \\
 0 & 0 & \frac{3}{2} & 0 & \frac{1}{6} \\
 0 & 0 & 0 & 1 & 0 \\
 0 & 0 & -\frac{15}{2} & 0 & -\frac{3}{2} \\
\end{array}
\right)\left(\begin{array}{c}{\cal L}_1^{\rm (AdS)} \\ {\cal L}_2^{\rm(AdS)} \\ {\cal L}_3^{\rm(AdS)} \\ {\cal L}_4^{\rm(AdS)} \\ {\cal L}_5^{\rm(AdS)}\end{array}\right)\, ,   \label{AdStoweymatre2}\ee 
and the inverse is
\be  \left(\begin{array}{c}{\cal L}_1^{\rm(AdS)} \\ {\cal L}_2^{\rm(AdS)} \\ {\cal L}_3^{\rm(AdS)} \\ {\cal L}_4^{\rm(AdS)} \\ {\cal L}_5^{\rm(AdS)}\end{array}\right)=\left(
\begin{array}{ccccc}
 1 & 0 & \frac{1}{8} & 0 & \frac{1}{24} \\
 0 & 1 & 0 & -\frac{1}{6} & 0 \\
 0 & 0 & \frac{3}{2} & 0 & \frac{1}{6} \\
 0 & 0 & 0 & 1 & 0 \\
 0 & 0 & -\frac{15}{2} & 0 & -\frac{3}{2} \\
\end{array}
\right) \left(\begin{array}{c}{\cal L}_1^{\rm (dS)} \\ {\cal L}_2^{\rm(dS)} \\ {\cal L}_3^{\rm(dS)} \\ {\cal L}_4^{\rm(dS)} \\ {\cal L}_5^{\rm(dS)}\end{array}\right)  \,. \label{AdStoweymatre2}\ee 
In terms of the coefficients $c_i$, $d_i$ of the general Lagrangians \eqref{genlagexdeadse}, \eqref{genlagexdedse} this map becomes the maps \eqref{ctodeqe1}, \eqref{ctodeqe2} which preserves the $S$-matrix elements that we computed.
These maps preserve the tadpole vanishing conditions \eqref{adstadpolee1}, \eqref{adstadpolee}, confirming that the map preserves the vacuum solution $\phi=0$.  They also preserve the kinetic normalization constant $Z$ in \eqref{adscanonirmee1}, \eqref{adscanonirmee} once the tadpole conditions are enforced, confirming that they are indeed perturbative field redefinitions.

\section{Discussion}

We have studied the DBI theory in which a flat brane is embedded into a standard AdS bulk or a dS bulk with 2 time dimensions.  These two bulks have the same isometry algebra, and in both cases the symmetry breaking pattern is the same.  Although the dS theory has a ``wrong sign'' DBI action due to the timelike extra dimension, we find that the two theories are nevertheless equivalent, related by an invertible field redefinition that we constructed by generalizing the ``AdS/CFT Equivalence Transformation'' of \cite{Bellucci:2002ji} (which we could thus also call a ``dS/CFT equivalence transformation'').  This confirms, despite the seemingly different physical setups, the expectation that an effective field theory is determined solely by its degrees of freedom and symmetry breaking pattern.  In particular, it confirms that there is a unique EFT for conformal symmetry breaking \cite{Hinterbichler:2022ids}, which we have seen can take the form of an (A)dS DBI theory, Weyl/dilaton theory, or any of the infinite other possible ways of parametrizing the coset.  The Weyl/dilaton theory has the advantage that the derivative power counting is manifest in the Lagrangians, whereas the (A)dS DBI parametrizations make explicit the geometric realization of the symmetry breaking, albeit in different geometries with different numbers of time dimensions.

We considered only cases where the brane is flat, so that we could study $S$-matrix elements of the theory, but a similar story should also go through for the DBI theories in which a curved brane is embedded into an (A)dS bulk \cite{Goon:2011qf,Goon:2011uw,Bonifacio:2018zex,Bonifacio:2021mrf}.  

Physics in a spacetime with more than one time dimension is often considered to be pathological for various reasons, such as a lack of causality, unitarity, stability, well-posedness, etc. (see e.g. \cite{Weinstein:2008aj,Foster:2010vt} for overviews).  However, even if such pathologies are unavoidable, it can be asked whether scenarios with extra dimensions might allow for timelike extra dimensions, with our four-dimensional universe containing only a single time direction, in such a way that the pathologies associated to the extra times are sequestered safely away from our universe.
Such higher dimensional scenarios with extra time dimensions have been considered before \cite{Dvali:1999hn,Matsuda:2000nk,Chaichian:2000az,Gogberashvili:2000yq,Gabadadze:2000bza,Iglesias:2000iz,Matsuda:2000cz,Berezhiani:2001bn,Quiros:2007db,Quiros:2007ym,Nortier:2020xms}, with varying answers as to whether the supposed pathologies manifest themselves in our four dimensions or remain confined in the extra dimensions.

The equivalence we studied here is an example where the on-brane physics of a fixed bulk spacetime with two time dimensions is indistinguishable from that of a spacetime with one time dimension, which implies there should be no pathology associated with the extra time dimension.  
One could argue that this is because the extra dimension is non-dynamical, and thus essentially a formal device for implementing the desired symmetries (as in e.g. 2T physics \cite{Bars:2000qm} or F-theory \cite{Vafa:1996xn}).  Thus our conclusion would likely change once the bulk becomes dynamical in any way, as for example in Dvali-Gabadadze-Porrati \cite{Dvali:2000hr} or Randall-Sundrum \cite{Randall:1999ee} type setups.  In these cases all the issues associated with two-time physics would presumably manifest themselves.  It may be interesting to study this in more detail, and it may also be interesting to study some intermediate cases where one keeps the bulk non-dynamical but gives dynamics to the brane geometry, either through the induced metric as in \cite{deRham:2010eu} or via an additional dynamical metric as in \cite{Gabadadze:2012tr}.

\vspace{-5pt}
\paragraph{Acknowledgments:}

KH would like to thank James Bonifacio, Austin Joyce and Diederik Roest for discussions and earlier collaboration during which the questions addressed here arose (see footnote 9 of \cite{Bonifacio:2021mrf}).  We acknowledge support from DOE grant DE-SC0009946.

\bibliographystyle{utphys}
\addcontentsline{toc}{section}{References}
\bibliography{samantapaper-arxiv}

\end{document}